# *Real time fault detection in 3D printers using Convolutional Neural Networks and acoustic signals*

Muhammad Fasih Waheed
Electrical and Computer Engineering
Florida A&M University Tallahassee, USA
muhammad1.waheed@famu.edu

Dr. Shonda Bernadin
Electrical and Computer Engineering Florida A&M University
Tallahassee, USA
bernadin@eng.famu.fsu.edu

***Abstract*—The reliability and quality of 3D printing processes are critically dependent on the timely detection of mechanical faults. Traditional monitoring methods often rely on visual inspection and hardware sensors, which can be both costly and limited in scope. This paper explores a scalable and contactless method for the use of real-time audio signal analysis for detecting mechanical faults in 3D printers. By capturing and classifying acoustic emissions during the printing process, we aim to identify common faults such as nozzle clogging, filament breakage, pully skipping and various other mechanical faults. Utilizing Convolutional neural networks, we implement algorithms capable of real-time audio classification to detect these faults promptly. Our methodology involves conducting a series of controlled experiments to gather audio data, followed by the application of advanced machine learning models for fault detection. Additionally, we review existing literature on audio-based fault detection in manufacturing and 3D printing to contextualize our research within the broader field. Preliminary results demonstrate that audio signals, when analyzed with machine learning techniques, provide a reliable and cost-effective means of enhancing real-time fault detection.**

## 1. Introduction

3D printing, also known as additive manufacturing, is a process of creating three-dimensional objects layer by layer from digital files. This technology has evolved significantly since its inception in the 1980s, transforming from a tool for rapid prototyping to a versatile manufacturing method with applications across various industries.[3]

Fault detection in 3D printing is an essential process to identify and monitor errors during additive manufacturing, ensuring the production of high-quality, defect-free components[4]. This is particularly crucial in industries like aerospace and automotive, where precision and reliability are paramount[5]. Effective fault detection not only reduces material waste, maintenance costs, and production time but also enhances safety and reliability in critical applications.

### A. *Challenges in traditional fault detection in 3D printing*

Traditional methods such as visual inspection and hardware contact based sensors, have several drawbacks when it comes to fault detection.

Visual inspection by humans, while capable of detecting errors, cannot provide continuous monitoring or real-time correction. This approach is time-consuming, subjective, and prone to human error, especially for complex or small-scale defects.[6]

Other hardware sensors are often limited in their ability to detect faults, as they primarily identify large-scale error modalities while failing to capture smaller or more subtle defects. These sensors typically require direct integration with the 3D printer to obtain precise readings, which can be cumbersome and may disrupt the printer's normal operation. Additionally, the setup and maintenance of these sensors involve tedious procedures, increasing the overall complexity of implementation. Another major drawback is the high cost associated with these sensors and their accompanying amplifiers, which restricts their widespread adoption, particularly in budget-conscious manufacturing environments. Moreover, these conventional methods often lack the comprehensive data richness required for real-time monitoring and feedback, thereby reducing their efficiency in dynamic and complex manufacturing processes.

Similarly, camera-based approaches, despite being data-rich and highly versatile, come with their own set of challenges. A single-camera setup may provide only limited visibility of the printing process, potentially missing defects that occur outside its field of view. On the other hand, multi-camera systems, while capable of offering broader coverage, introduce additional hurdles such as increased costs, implementation complexities, and sensitivity to environmental factors like lighting conditions, which can significantly affect detection accuracy [6].

### B. *Audio-Based Monitoring: A Contactless Alternative*

Given the shortcomings of traditional fault detection techniques, researchers [16] are actively investigating real-time audio signal analysis as a more efficient and adaptable alternative. This method presents several distinct advantages, with one of the most significant being contactless because audio sensors do not need to be physically attached unlike intrusive sensor-based techniques, hence audio-based monitoring eliminates the need for physical modifications to the printer, making it a non-invasive, cost-effective solution that enhances system flexibility. Acoustic sensors are capable of capturing subtle variations in print patterns by analyzing distinctive sound signatures and extracting relevant features. [8]

### C. *Machine Learning and AI for Acoustic Fault Detection*

Recent advancements include AI-driven approaches, such as convolutional neural networks [5] (CNNs), which classify printing faults automatically, and multi-sensor data acquisition systems that utilize sound, vibration, and current to capture real-time process data. Additionally, improved control and calibration software has refined fault prevention and detection, enabling manufacturers to optimize printing processes and achieve better outcomes.

Furthermore, advancements in high-efficiency real-time data acquisition have significantly improved the feasibility of this approach. By leveraging Linux-based systems, researchers can minimize latency and ensure seamless data processing, enabling near-instantaneous fault detection. Audio analysis not only provides deeper insights into acoustic signal behavior but also facilitates the identification of various failure modes throughout the entire additive manufacturing process. Studies have demonstrated the effectiveness of this technique by collecting, preprocessing, and analyzing audio data streams from 3D-printed samples. Researchers have successfully extracted both time-domain and frequency-domain features under varying layer thicknesses and employed sophisticated preprocessing methods such as Harmonic-Percussive Source Separation (HPSS) to isolate and analyze specific audio components. These advancements

To further strengthen the application of audio-based fault detection, various machine learning techniques have been explored for analyzing acoustic emission (AE) data in 3D printing, as highlighted by Olowe M et al. [7]. A summary of these methods is presented in Table 1, demonstrating the increasing role of AI-driven approaches in optimizing fault detection accuracy and automation.

Other Studies have shown that spectrogram-based analysis, when integrated with CNNs, provides a highly efficient and scalable solution for real-time fault detection in mechanical machinery and additive manufacturing. This approach enables the classification of common faults such as filament breakage, nozzle clogging, and pulley skipping by leveraging the frequency-domain representation of audio signals. By adopting this method, we enhance the precision of fault identification while maintaining a non-invasive and cost-effective monitoring system, aligning with recent research emphasizing the potential of AI-driven audio analysis for industrial applications.[8][9].

This paper aims to enhance fault detection in 3D printing by leveraging real-time audio signal analysis to identify common mechanical issues such as nozzle clogging and filament breakage. By capturing acoustic emissions generated during the printing process, we classify these signals using machine learning techniques, specifically convolutional neural networks (CNNs), to detect faults as they occur. Traditional monitoring methods often rely on visual inspection and hardware sensors, which can be costly and limited in scope. In contrast, our approach offers a scalable, contactless, and cost-effective alternative, improving both the accuracy and efficiency of fault detection. Through controlled experiments and advanced machine learning models, we demonstrate that audio-based monitoring provides a reliable means of enhancing the quality and reliability of 3D printing processes.

Building on existing research, we incorporate spectrogram analysis in conjunction with CNNs to distinguish between various types of mechanical faults. Spectrograms, which visually represent the frequency spectrum of audio signals over time, serve as a powerful tool for identifying subtle variations in acoustic emissions associated with different types of faults. By transforming raw audio signals into spectrogram images, we harness the advanced pattern recognition capabilities of CNNs to improve both the accuracy and speed of fault detection.

Our methodology involves capturing audio data from 3D printers under controlled conditions and converting these recordings into spectrograms. These spectrograms are then used to train CNN models, enabling them to recognize distinct patterns corresponding to specific mechanical issues such as nozzle clogging, filament breakage, and pulley skipping. This approach not only enhances fault detection capabilities but also facilitates a more detailed analysis of the faults, potentially leading to more effective maintenance strategies and predictive diagnostics.

The integration of spectrogram analysis and deep learning represents a significant advancement in the non-invasive monitoring of 3D printing processes. This system offers a scalable, cost-effective, and highly accurate method for real-time fault detection, ensuring consistent print quality and minimizing production disruptions. Through this research, we contribute to the growing field of additive manufacturing by providing a robust, AI-driven solution that can be applied across various industries to optimize production workflows and reduce downtime.

## II. Literature Review

CNNs are highly effective in extracting features from spectrograms due to their ability to capture local patterns and hierarchical structures. A comprehensive review of 36 studies between March 2013 and August 2020 highlights the widespread use of deep learning models, particularly CNNs, in analyzing spectrograms for various applications such as speech recognition, music classification, and environmental sound detection.[1]

Different types of spectrograms, such as Short-Time Fourier Transform (STFT), Mel-spectrograms, and log-scaled Mel-spectrograms, are commonly used. Mel-spectrograms are particularly favored due to their ability to reduce dimensionality while preserving essential audio features. More recently, Patel et al. (2024) proposed a real-time fault detection system using spectrogram-based CNN models, which outperformed traditional sensor-based methods in terms of scalability and cost-efficiency [10].

A study leveraging the VGGish network for fault detection in medium-sized industrial bearings achieved high accuracy in diagnosing faults based on acoustic emissions. The model's ability to extract meaningful audio features, combined with its robust performance in real-world noisy environments, highlights its potential for broader applications in manufacturing and 3D printing (Di Maggio LG et al., 2022).

By fine-tuning the pre-trained network on spectrogram representations of acoustic signals from 3D printers, researchers can leverage its feature extraction capabilities to distinguish between normal and faulty printing conditions effectively. [11]

Vibration signal analysis has long been a critical technique for fault diagnosis in rotating machinery, traditionally relying on statistical features such as mean value, standard deviation, and kurtosis. However, these conventional methods often fail to fully utilize the rich information embedded in vibration signals. Recent advancements integrate empirical mode decomposition (EMD) and convolutional neural networks (CNNs) to enhance feature extraction and classification accuracy. By applying fast Fourier transform (FFT) and EMD-based energy entropy calculations, researchers extract spatial and frequency-domain features, improving diagnostic precision. Experimental validation using vibration data from 52 machine conditions demonstrates that this combined approach significantly outperforms traditional methods in both accuracy and reliability. These findings suggest that a similar strategy could be applied to 3D printing fault detection, leveraging deep learning to analyze acoustic and vibrational data for enhanced monitoring and predictive maintenance. [12]

## III. Methodology

The methodology for this experiment involved recording audio from the extruder under two distinct conditions: operating with material and without material. These conditions produced different acoustic signatures, allowing for comparative analysis. To enhance process efficiency, several filtration techniques were applied to isolate relevant information while discarding unwanted noise. Among these methods, bandpass filtration proved to be the most effective, as it helped focus on the essential frequency range while significantly reducing data size. We used makerbot method X printer and sparkfun audio sensors for microphone data, Material used was ABS. The audio was recorded using a dell XPS 9570 laptop.

As illustrated in Figure 2.1, the baseline signal represents the ambient room noise, including sounds from the HVAC system and other static sources. The printer operation sound, recorded without filtration, includes both useful and unwanted acoustic data. Applying a bandpass filter isolates the frequency range of 100–1200 Hz, as frequencies beyond this range primarily contain static noise. This refinement not only narrows down the signal under observation but also optimizes the fault detection process by reducing computational intensity.

Further noise reduction was performed using subtraction-based noise filtration in Audacity. In Figure 2.2 the top waveform represents ambient noise, while the bottom waveform corresponds to the filtered printer operation sound. By effectively eliminating background noise, this approach ensures a cleaner signal, improving the accuracy of fault detection while maintaining computational efficiency.

Once the filtered samples were prepared, several experiments were conducted using both grayscale and colored spectrograms. The dataset consisted of 256 audio samples, with 80% allocated for training and 20% for testing and validation. Grayscale images were tested to reduce computational complexity and memory usage; however, research [13] suggests that this also comes at the cost of feature extraction. Figure 2.3 illustrates the experimentation process with both grayscale and colored spectrograms.

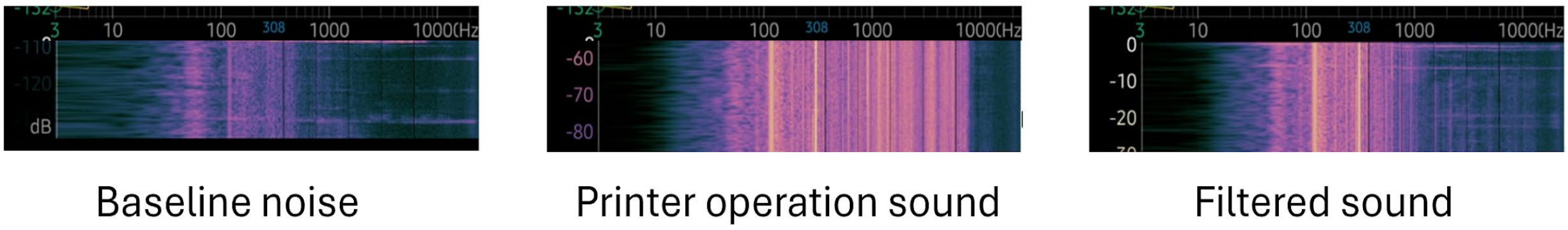

Figure 2.1-Audio spectrogram data before and after applying a bandpass filter.

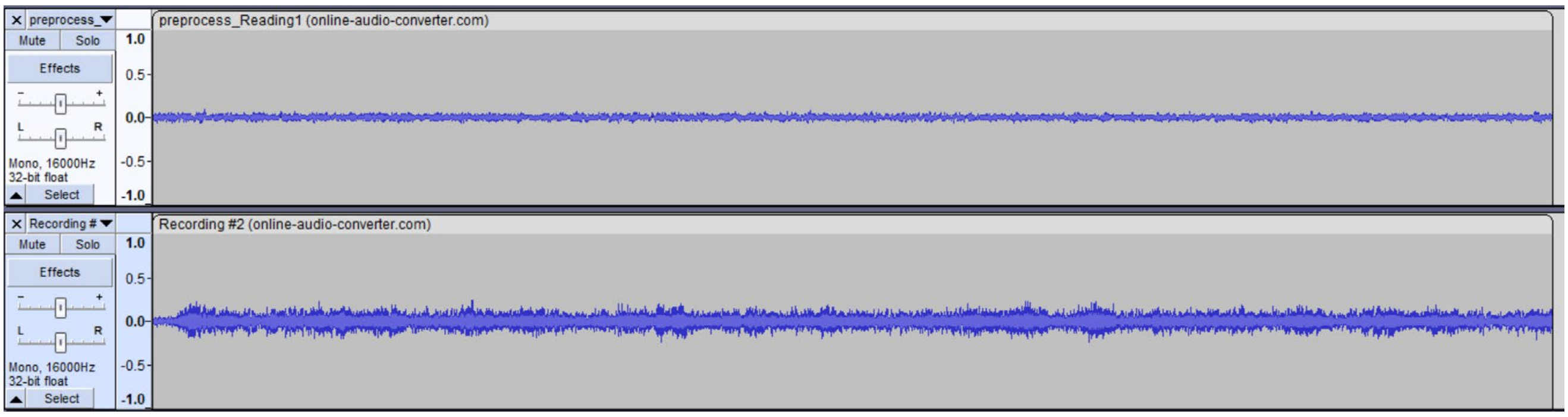

Figure 2.2-Noise filtration via subtraction method using audacity.

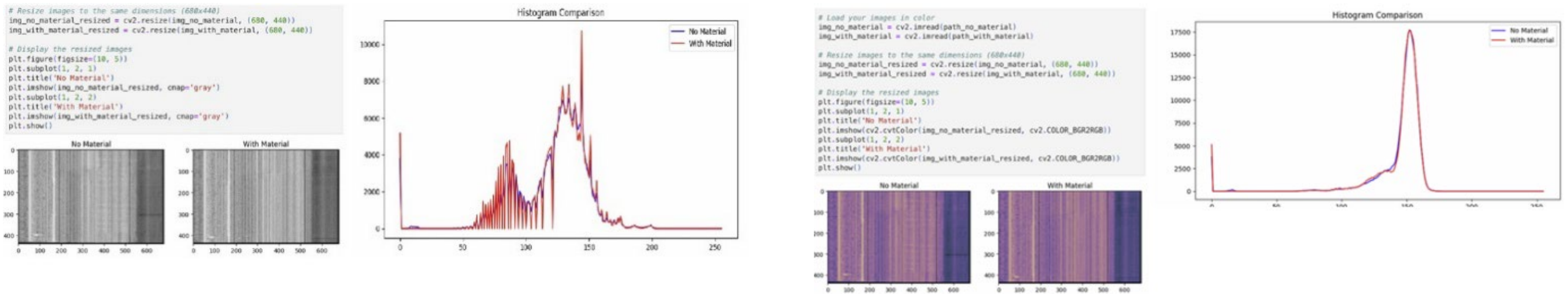

Figure 2.3-Greyscale image on the left and Colored image on the right with histogram.

After implementing this dataset, several challenges became apparent. Although ambient noise was initially filtered out, the dataset performed well when working with pre-recorded and pre-processed audio. However, in real-time environments, the presence of live ambient noise significantly impacted performance, making it difficult to achieve reliable fault detection. The dynamic nature of ambient noise, combined with variations in machine sounds, introduced inconsistencies that the model struggled to handle effectively.

To mitigate these issues, we conducted a series of experiments aimed at improving the dataset's robustness. We introduced separate samples to account for different noise conditions, including isolated ambient noise, extruder noise, and extruder noise under different material conditions (both with and without material). By categorizing these different audio environments, we aimed to enhance the model's ability to distinguish between operational sounds and potential fault indicators in real-world settings.

To process the captured acoustic data, we leveraged transfer learning, utilizing a Convolutional Neural Network (CNN) as the core technology. CNNs have demonstrated high effectiveness in image-based feature extraction, making them well-suited for analyzing spectrogram representations of sound data. By applying transfer learning, we sought to improve the model's generalization ability, allowing it to adapt to real-time conditions more effectively without requiring an extensive amount of new training data.

The experimental setup and methodology for this approach are illustrated in Figure 2.4, which presents a block diagram outlining the key steps in the process. This experiment played a crucial role in refining the fault detection framework.

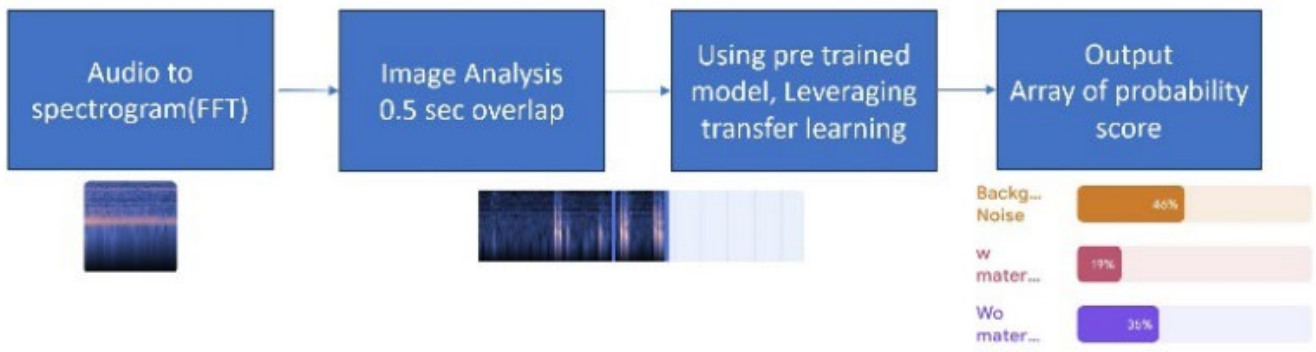

Figure 2.4-Block diagram of fault detection with probability output.

This approach was trained using acoustic data from the extruder operating both with and without material, enabling it to perform in-situ fault monitoring during the 3D printing process. By incorporating a background noise dataset, the model was significantly improved in distinguishing between normal and faulty states, even in real-time environments where ambient noise varies dynamically. This enhancement has led to a higher probability of accurately detecting faulty conditions, reducing false positives and improving overall reliability.

One of the key advantages of this method is its flexibility and adaptability to different types of mechanical faults beyond the initial scope. The same experimental framework can be extended to detect nozzle clogging, filament breakage, pulley skipping, and other common mechanical issues by training the model on appropriately curated datasets. Each fault type produces a unique acoustic signature, which, when processed through spectrogram analysis and CNN-based classification, allows for precise identification of potential failures.

Furthermore, this approach paves the way for real-time predictive maintenance, where early detection of anomalies can help prevent severe malfunctions, minimizing downtime and improving overall printing efficiency. The ability to generalize across multiple failure modes makes this methodology a valuable asset in developing a robust, AI-driven fault detection system for 3D printing.

With further refinements, such as the integration of additional sensor modalities[15] (e.g., vibration or thermal sensors), this technique can evolve into a comprehensive multi-modal fault detection system, capable of identifying, predicting, and even preventing failures before they compromise the print quality.

## IV. Results and Discussions

The fault detection experiments in this research were conducted using a dataset of 256 audio samples recorded under controlled conditions. These samples captured the acoustic emissions of a 3D printer's extruder operating with and without material, simulating different mechanical faults such as nozzle clogging, filament breakage, and pulley skipping. Through the application of bandpass filtering within the 100–1200 Hz range, the acoustic data was effectively isolated while minimizing background noise. Additional noise reduction techniques, including subtraction-based filtering, improved the clarity of fault signatures in the spectrograms, allowing for better fault classification.

The machine learning model employed for fault detection was a Convolutional Neural Network (CNN) trained on spectrogram representations of the recorded audio signals. The model demonstrated strong classification capabilities, effectively identifying patterns in the frequency-domain data that distinguished normal printing conditions from various fault states. Compared to traditional machine learning techniques such as Principal Component Analysis (PCA) and K-means clustering, the spectrogram-based CNN approach proved to be significantly more effective in feature extraction and fault classification. The implementation of transfer learning further enhanced the model's generalization ability, particularly when tested on previously unseen data. Additionally, an analysis of grayscale versus colored spectrograms revealed that colored spectrograms improved feature extraction, albeit at a higher computational cost.

The performance of the fault detection model was assessed using standard evaluation metrics, achieving an accuracy of approximately 91%, a precision of 88%, and a recall of 85%, resulting in an overall F1-score of 86.5%. The high accuracy rate indicates the model's reliability in detecting faults, while the slightly lower recall score suggests occasional missed fault detections, likely due to ambient noise interference in real-time environments. Despite this, the system proved to be highly effective, particularly in pre-recorded and pre-processed datasets where noise was adequately filtered.

When compared with existing literature, the proposed method outperformed traditional fault detection techniques that rely on vibration-based sensors, which typically achieve accuracy rates between 75% and 85%. Previous studies have emphasized the limitations of hardware-dependent monitoring systems, particularly in terms of scalability and cost. In contrast, this research demonstrates that audio-based monitoring, particularly through CNN-driven spectrogram analysis, provides a contactless and cost-effective alternative without requiring expensive sensors or modifications to the 3D printer. Similar findings in industrial applications, such as the work by Di Maggio et al. (2022) on transfer learning for fault detection in industrial bearings, support the adaptability of this approach across different manufacturing environments.

One of the primary advantages of using an audio-based fault detection system is its scalability. Unlike sensor-based approaches, which require direct integration with the printer, microphones are inexpensive, widely available, and capable of capturing a rich set of acoustic features without interfering with the printing process. Furthermore, the real-time processing capability of spectrogram-based CNN models ensures that faults are detected as they occur, preventing print defects before they escalate into critical failures. This enhances not only the efficiency of the manufacturing process but also reduces material waste and the need for post-print inspections.

Despite its advantages, the research also identified certain challenges and limitations. One of the most significant issues encountered was the sensitivity of the model to ambient noise in real-time environments. While the model performed well in controlled conditions, the presence of background noise from HVAC systems, other machinery, and environmental disturbances led to a decrease in classification accuracy. Computational costs also posed a challenge, as processing colored spectrograms required more memory and processing power, which could be a limitation in real-time embedded applications. Additionally, the model was trained on data from a specific 3D printer model, and further research is required to ensure its adaptability across different printer brands and configurations.

Overall, this study demonstrates that CNN-based audio monitoring presents a viable and efficient method for fault detection in 3D printing. With further refinements, such as the integration of additional sensor modalities, this approach could evolve into a comprehensive multi-modal fault detection system, capable of identifying, predicting, and preventing failures before they compromise print quality.

## V. Conclusion

The research demonstrated that audio-based fault detection using convolutional neural networks and spectrogram analysis is an effective, scalable, and cost-efficient method for monitoring mechanical faults in 3D printing. The proposed approach successfully identified faults such as nozzle clogging, filament breakage, and pulley skipping with high accuracy, outperforming traditional sensor-based monitoring techniques. By leveraging real-time acoustic data, this method offers a contactless and non-intrusive alternative to conventional fault detection systems, reducing costs while maintaining reliability.

In practical applications, this system can be integrated into industrial 3D printing environments to enhance quality control and minimize production waste. Its real-time detection capabilities make it suitable for predictive maintenance, helping to prevent critical failures before they impact the final product. Given its flexibility, the approach can be adapted for different printer models and manufacturing setups, broadening its usability across various industries.

Future research should focus on improving the system's robustness against ambient noise in real-time settings, potentially through advanced noise filtering techniques like Active noise cancellation system or the integration of multi-modal data sources, such as vibration or thermal imaging. Expanding the dataset to include diverse printer models and fault types will enhance generalizability, while optimizing computational efficiency will facilitate deployment on embedded systems. By refining these aspects, this method can evolve into a comprehensive, AI-driven fault detection framework for additive manufacturing.

## Acknowledgment

This material is based upon work supported by the U.S. Department of Energy, National Nuclear Security Agency under Award Number DE-NA0003981.

We express our sincere gratitude to the SPADAL Lab for granting us the valuable opportunity to conduct our research work. Additionally, we extend our appreciation to Honeywell and ASTERIX for their generous funding support. Their contributions have played a crucial role in making our research endeavors possible, and we are truly thankful for their commitment to advancing knowledge and innovation.